\documentclass[journal=jacsat,manuscript=article]{achemso}

\usepackage{hyperref}
\usepackage[version=3]{mhchem} 
\usepackage{booktabs}
\usepackage{siunitx}

\usepackage[table]{xcolor}
\usepackage{multirow}
\usepackage{array}
\usepackage{tikz}

\definecolor{lightgreen}{RGB}{180,230,180}

\author{Franz Görlich}
\affiliation[TUM]{Professorship of Multiscale Modeling of Fluid Materials, Department
of Engineering Physics and Computation, TUM School of Engineering
and Design, Technical University of Munich, Germany}

\author{Julija Zavadlav}
\email{julija.zavadlav@tum.de}
\affiliation[TUM]{Professorship of Multiscale Modeling of Fluid Materials, Department
of Engineering Physics and Computation, TUM School of Engineering
and Design, Technical University of Munich, Germany}
\alsoaffiliation[AMC]{Atomistic Modeling Center (AMC), Munich Data Science Institute
(MDSI), Technical University of Munich, Germany}

\title{Mapping Still Matters: Coarse-Graining with Machine Learning Potentials}

\begin{document}


\begin{abstract}
Coarse-grained (CG) modeling enables molecular simulations to reach time and length scales inaccessible to fully atomistic methods. For classical CG models, the choice of mapping, that is, how atoms are grouped into CG sites, is a major determinant of accuracy and transferability. At the same time, the emergence of machine learning potentials (MLPs) offers new opportunities to build CG models that can in principle learn the true potential of the mean force for any mapping. In this work, we systematically investigate how the choice of mapping influences the representations learned by equivariant MLPs by studying liquid hexane, amino acids, and polyalanine. We find that when the length scales of bonded and nonbonded interactions overlap, unphysical bond permutations can occur. We also demonstrate that correctly encoding species and maintaining stereochemistry are crucial, as neglecting either introduces unphysical symmetries. Our findings provide practical guidance for selecting CG mappings compatible with modern architectures and guide the development of transferable CG models.
 
\end{abstract}

\section{Introduction}

Molecular dynamics (MD) simulations have become an indispensable tool in chemistry, biology, and materials science, offering atomistic insights into the behavior of complex systems. However, a persistent challenge is the vast range of time and length scales governing molecular phenomena. Many processes, such as protein folding or polymer dynamics, occur in microseconds or longer, far exceeding the time scales accessible to conventional all-atom simulations \cite{noe2020machine, kmiecik2016coarse}. To bridge this gap, coarse-graining (CG) methods can be used. CG models simplify the system by grouping atoms into fewer interaction sites, or "beads", thus reducing the effective degrees of freedom. This allows for significantly larger and longer simulations. The central goal of CG models is to preserve the essential structural and thermodynamic properties of the original atomistic system \cite{noid2013perspective, kmiecik2016coarse, noid2023perspective}.

When developing a coarse-grained (CG) model, two key questions typically arise: 1. What functional form should be used to represent the CG potential and 2. What mapping scheme should be adopted to define how atoms are grouped into beads? In atomistic systems, only the choice of a potential functional form is relevant, and extensive work has focused on improving its accuracy. In the last decade, attention has been mostly focused towards machine learning potentials (MLPs).\cite{kocer2022neural, thiemann2024introduction} Their success is based on a paradigm shift that also occurred in other domains, where MLPs learn interactions directly from data, compared to hand-crafted interaction terms of classical potentials.\cite{behler2021machine, kocer2022neural}

Many modern MLPs incorporate physical symmetries into the model architecture. In particular, E(3)-equivariant neural networks enforce rotational, translational, and reflection symmetries inherent to molecular systems. Architectures such as NequIP\cite{batzner20223}, Allegro\cite{musaelian2023learning} and MACE\cite{MACE} that built on these principles have demonstrated exceptional accuracy and data efficiency.
The application of equivariant MLPs to coarse-grained modeling is a natural and promising extension.\cite{wilson2023anisotropic,loose2023coarse, mirarchi2024amaro} An early study of CG liquid water indicates that equivariant MLPs can drastically reduce the amount of training data needed, performing reasonable with as little as a single reference frame.\cite{loose2023coarse} While single-bead mappings are trivial, it is unclear how more general mapping choices affect the learned representation.\cite{loose2023coarse} In practice, many existing CG-MLPs still rely on priors, i.e. classical energy terms, such as harmonic bond or angle potentials, to maintain molecular connectivity and physical accuracy.\cite{wang2019machine, thaler2022deep, durumeric2023machine, charron2025navigating, ge2025coarse} This practice reintroduces the manual heuristics that atomistic MLPs were designed to avoid.\cite{behler2016perspective, unke2021machine}

In this study, we investigate how modern E(3)-equivariant machine learning potentials can be applied to coarse-graining without the use of any priors. Using the MACE architecture as a representative model\cite{MACE}, we perform a systematic study to evaluate how the choice of mapping, species encoding and model parameters influence the stability and learned representation of the model. We compare these results with a classical CG potential to assess whether the learned models provide a more accurate description of coarse-grained energy landscapes. We evaluated three distinct systems of increasing complexity: liquid hexane, single amino acids, and a 15-mer polyalanine. We validate key findings by additionally testing them with NequIP \cite{batzner20223}.

Our results show that the mapping choice can significantly influence the representation that equivariant MLPs learn. In low resolution mappings of amino acids, we observe symmetries in the free energy surface (FES). We identified that these are caused by enantiomerization or ambiguous species encoding. We show that such symmetries propagate from symmetric dihedrals in single amino acids to incorrect secondary structure formation in larger peptides. Finally, we show that bond permutations can occur when bonded and nonbonded length scales overlap, such as in the two-site hexane or $C_\alpha$ polyalanine model. Together, these findings highlight that while equivariant MLPs offer remarkable flexibility and data efficiency,\cite{loose2023coarse, wang2019machine} the preservation of a faithful representation critically depends on the CG mapping.

\section{Methods}

\subsection{Coarse-graining}
In coarse-graining (CG), the degrees of freedom of an atomistic system are reduced by grouping different atoms into beads. Here we limit the study to mappings, in which each atom at most contributes to one CG bead. We use the center-of-mass (COM) mapping to derive CG positions and forces. Other, more general mapping choices\cite{babadi2006coarse, gay1981modification, Voth2008-zb}, as well as optimal mappings\cite{diggins2018optimal, yang2023slicing, wang2019coarse}, have also been explored, but remain less prevalent in practice. In the COM mapping, the CG positions $\textbf{R} \in \mathcal{R}^{3N}$ are derived as a linear mapping $\textbf{M}$ of the atomistic positions $\textbf{r} \in \mathcal{R}^{3n}$, where $N <n$:

\begin{equation}
    \textbf{R}_I = \textbf{M}_I(\textbf{r}) =  \sum_{i \in \mathcal{S}_I}w_{Ii}\textbf{r}_i
\end{equation}

Here, $\mathcal{S}_I$ denotes the set of atom indices that contribute to the CG bead $I$. Each atom can contribute only to one CG bead. In the COM mapping, the contribution of each atom $w_{Ii}$ to the CG bead $I$ is equivalent to its contribution to the total mass:

\begin{equation}
w_{I_i} =
\begin{cases}
\frac{m_i}{\sum_{j \in \mathcal{S}_I} m_j} \quad   & \text{if } i \in \mathcal{S}_I \\
0 & \text{otherwise}
\end{cases}
\end{equation}

The weights are normalized such that $\sum_{i \in \mathcal{S}_I} w_{Ii} = 1$.\cite{noid2008multiscale} In the COM mapping, the effective force acting on a coarse-grained bead $\text{f}_I$ is simply the sum of the atomistic forces contributing \cite{noid2008multiscale, john2017many}

\begin{equation}
    \text{f}_I(\textbf{r}) = \sum_{i \in \mathcal{S}_I} f_i(\textbf{r})
\end{equation}

CG models are parametrized in a "top-down" or "bottom-up" manner. Bottom-up methods\cite{kock2008growth, ercolessi1994interatomic, shell2008relative} use a more detailed reference, for example an atomistic simulation or quantum-mechanics, to derive the coarse-grained potential, while top-down approaches\cite{white2015designing, reith2003deriving} aim to reproduce macroscopic observables, such as experimental measurements.\cite{jin2022bottom} Many models also implement a hybrid approach, for example the MARTINI forcefield\cite{souza2021martini}, which parametrizes nonbonded interactions via experimental partition coefficients, while bonded terms are derived from classical atomistic simulations.\cite{ thaler2021learning} Here we employ force matching, a bottom-up method, to derive the CG potentials.

\subsection{Force Matching}
Force matching, first introduced by Izvekov and Voth as the multiscale coarse graining (MS-CG) method\cite{izvekov2005multiscale}, optimizes the effective force acting on the CG bead $\text{f}_I$ by minimizing the least-squared force residual $\chi^2$ between the predicted CG and the mapped atomistic forces \cite{jin2022bottom, noid2024rigorous}.

\begin{equation} \label{fm_loss}
    \chi^2(\theta) = \ 
    \left\langle \frac{1}{3N}\sum_{I=1}^N | \text{f}_I(\textbf{r}) + \nabla_I U_{CG}(\textbf{M}_I(\textbf{r}),\theta)  |^2 \right\rangle
\end{equation}

Here $ \text{F}_I(\textbf{R},\theta) = -\nabla_I U_{CG}(\textbf{R},\theta)$ are the predicted forces as a function of the model parameters $\theta$.

\subsection{Coarse-grained Potential}
\paragraph{Classical Potential} 
We employ the \texttt{VOTCA} framework \cite{ruhle2009versatile} to derive classical coarse-grained potentials $U_{Classical}(\mathbf{R};\theta)$. The general form is a sum of bonded (bonds, angles, dihedrals) and nonbonded interactions, which are parametrized in the form of splines.
\begin{equation}
    U_{Classical}(\mathbf{R,\theta})= \sum_{bond}U_{bond} + \sum_{angle}U_{angle} + \sum_{dihedral}U_{dihedrals} + \sum_{nonbonded}U_{nonbonded} 
\end{equation}
We provide a detailed overview of the parametrized interactions and hyperparameters used in the Supporting Information.

\paragraph{Machine Learning Potential}
To parameterize an equivariant MLP $U_{MLP}(\mathbf{R};\theta)$, we employ the MACE architecture\cite{MACE}. MACE combines the idea of atomic cluster expansion\cite{drautz2019atomic} with message-passing neural networks (MPNNs)\cite{gilmer2017neural}. In MACE, the messages passed between nodes, in our case CG beads, are expanded using the idea of a hierarchical body order expansion over neighbors $J$:
\begin{equation}\label{MACE_message_corr}
\mathbf{m}_I^{(l)} = \sum_{J_1} u_1(\sigma_I^{(l)}, \sigma_{J_1}^{(l)}) + \dots + \sum_{J_1, \ldots, J_\nu} u_\nu(\sigma_I^{(l)}, \sigma_{J_1}^{(l)}, \ldots, \sigma_{J_\nu}^{(l)}),
\end{equation}
where $u_{1..\nu}$ are learnable functions and $\sigma_I^{(l)}$ is the state of CG bead $I$ in layer $l$. The total body order of the model depends on the number of layers $L$ and correlation order $\nu$.  

We use the \texttt{chemtrain}\cite{fuchs2025chemtrain, fuchs2025chemtraindeploy} framework to train the MACE potentials. The training process directly adjusts the model parameters based on equation \ref{fm_loss} via backpropagation. A detailed overview of the setup and model hyperparameters can be found in the Supporting Information.

\subsection{Simulations}
\paragraph{Reference Simulations}
Atomistic reference data was generated using GROMACS~\cite{gromacs} in the canonical ensemble (NVT) at 300 K. For liquid hexane, we followed the protocol of Ruehle et al.~\cite{ruhle2011hybrid}, employing the OPLS-AA force field~\cite{robertson2015improved} with a system size of 100 molecules. For peptide systems, we employed the AMBER ff99SB-ILDN force field with the TIP3P water model~\cite{lindorff2010improved}. All peptide structures were capped with N-terminal acetyl (ACE) and C-terminal N-methyl amide (NME) groups. Production runs were performed for 100 ns (liquid hexane) and 500 ns (peptides). The configurations were sampled uniformly to produce 500,000 training frames. Detailed simulation parameters and equilibration protocols are provided in the Supporting Information. 

\paragraph{CG Simulations}
Coarse-grained simulations were performed in the NVT ensemble at 300 K. Classical CG simulations were run in GROMACS using stochastic dynamics (SD). MLP simulations were performed using the JAX-MD engine within the \texttt{chemtrain} framework.\cite{fuchs2025chemtrain} A time integration step of 2 fs was used for all CG models. For the implicit solvent baseline of capped amino acids, a timestep of 0.5 fs was employed. For capped alanine, we also perform stability tests in a microcanonical (NVE) ensemble. Detailed simulation parameters are provided in the Supporting Information.

\subsection{Structural Analysis of Coarse-grained Simulations}

\paragraph{Radial Distribution Function}
To describe nonbonded interactions in the liquid hexane model, we employ radial distribution functions (RDFs). The RDF $g_{A-B}(r)$ captures the probability of finding a particle of species $A$ at a distance $r$ from a particle of species $B$. It is defined as the ratio of the local density of $A$ at a distance $r$ from $B$ to the bulk density of $A$, $\rho_A$:
\begin{equation}
    g_{A-B}(r) = \frac{dN_{A-B}(r)}{4\pi r^2\rho_A dr}
\end{equation}

where $dN_{A-B}(r)$ is the average density of $A$ particles found in a spherical shell of thickness $dr$ at a distance $r$ from a central particle $B$, and $\rho_A = N_A/V$ is the bulk number density of species $A$. Bonded atoms/CG beads are excluded for the RDF calculation.\cite{freud2020, thompson2022lammps}

\paragraph{Bonded parameters}
For each liquid hexane mapping, we show the bonded order parameter that embodies the most information: the dihedral angle $\phi_{A-B-B-A}$ for the four-site mapping, the angle $\theta_{A-B-A}$ for the three-site mapping, and the bond distance $b_{A-A}$ for the two-site mapping. To compare the conformational space of single capped amino acids mappings, we evaluated the two backbone dihedrals $\phi_{C_{ACE}-N-C_{\alpha}-C}$ and $\psi_{N-C_{\alpha}-C-N_{NME}}$. We compare 2D representations of the free energy surface (FES) of both dihedrals. The FES was calculated from the joint probability density $P(\phi, \psi)$ using the Boltzmann inversion relation $F(\phi, \psi) = -k_B T \ln P(\phi, \psi)$.

\paragraph{Polyalanine Helix}
To analyze the helix formation of the polyalanine peptide, we analyzed different order parameters based on the positions of the $C_\alpha$ atoms. We adapted the fractional helix content or helicity from Rudzinski et al \cite{rudzinski2015bottom}. This metric iterates over all $N_{hel}$ pairs of $C_\alpha$ atoms separated by three bonds, corresponding to the hydrogen-bonding pattern charactersitic of $\alpha$-helices. For each pair, the distance $d_{ij}$ is computed and scored by how close it's to an optimal distance $d_0 = 0.5$ nm with standard deviation $\sigma^2 = 0.02$ nm$^2$. $Q_{hel}$ ranges from 0, completely unstructured (coil or unfolded), to 1, a perfect alpha helix. 

\begin{equation}\label{helicity_eq}
    Q_{hel}(\mathbf{R}=\mathbf{R_{C_\alpha}}) =\frac{1}{N_{hel}}\sum_{i-j=3}\exp \!\left(-1/2\sigma^2(d_{ij} -d_o)^2\right)
\end{equation}

To differentiate left- from right-handed helices, we used the approach by Sidorova et al\cite{sidorova2021protein}. This method first constructs vectors between neighboring $C_\alpha$ atoms and then sums the mixed product of consecutive triplets of vectors:

\begin{equation}\label{handedness_eq}
    \chi_{hel}(\mathbf{R}=\mathbf{R_{C_\alpha}}) = \frac{1}{N_{hel}}\sum_{i=1}^{N_{C\alpha}-3}v_i \cdot (v_{i+1}\times v_{i+2})
\end{equation}

The sign of $\chi_{hel}$ determines the handedness or chirality sign of the helix. A positive $\chi_{hel}$ indicates a right-handed helix, while negative values indicate a left-handed helix.\cite{sidorova2021protein} The free energy $F(\chi_{hel},Q_{hel})$ was obtained through Boltzmann inversion from the joint probability density. To evaluate the helix propensity of the different mappings, we ranked all reference simulation frames according to $|\chi_{\mathrm{hel}} \cdot Q_{\mathrm{hel}}|$, selected the 100 frames with the lowest values, and performed 5~ns simulations starting from each configuration.

\subsection{Simulation Stability}\label{sec:sim_stability}
To assess simulation stability, we applied two different methods.  
For bulk systems (liquid hexane), we consider the conservation of thermal energy, \( k_BT \). Specifically, we check whether \( k_BT < 5 \) kJ/mol, where $k_B$ is the Boltzmann constant and $T=300$ K. If a particular frame exceeds this threshold, that frame and all subsequent frames are excluded. For the capped amino acids and polyalanine system, we adapt the distance-based stability criterion of Fu et al.\cite{fu2022forces}, by checking whether the distance between the bonded atoms deviates by more than 0.05 nm from the equilibrium bond length. The equilibrium bond length is calculated as the mean of the reference simulation for all matching bond types and frames.

We chose two different strategies, as a distance-based metric is not applicable to all mappings of bulk hexane: Bond distances might change drastically, while thermal energy is preserved (see Results). In the case of polyalanine, the distances might also change due to unphysical bond switching; however, the definition of the polyalanine order parameters depends on the correct order of the $C_\alpha$ beads. The list of equilibrium lengths as well as the results of the simulation stability analysis is provided in the Supporting Information.

\subsection{Well-tempered Metadynamics}  
To assess the free energy barrier for enantiomerization, we employ well-tempered metadynamics (WTMetaD).\cite{barducci2008well} Unlike standard metadynamics, which builds a bias potential by adding Gaussians of constant height,\cite{Laio2002} WTMetaD ensures convergence by rescaling the Gaussian height according to the accumulated bias. The history-dependent bias potential $V(\mathbf{s}, t)$ acting on a set of collective variables in the CG space $\mathbf{s}(\mathbf{R})$ is given by:

\begin{equation}
    V(\mathbf{s}, t) = \sum_{t' < t} 
    h \exp \!\left( - \frac{V(\mathbf{s}(t'), t')}{k_B T (\gamma - 1)} \right)
    \exp \!\left( - \sum_{i=1}^{d} \frac{(s_i - s_i(t'))^2}{2\sigma_i^2} \right),
\end{equation}

Here $h$ is the initial Gaussian height, $\sigma_i$ is the width of the $i$-th collective variable, and $\gamma = T^*/T$ is the bias factor. This corresponds to sampling the collective variables at an effective temperature $T^* = \gamma T$, which scales the free energy barriers by $1/\gamma$.

\section{Results and Discussion}
\subsection{Liquid Hexane}

\begin{figure}[H]
    \centering
    \includegraphics[width=16cm]{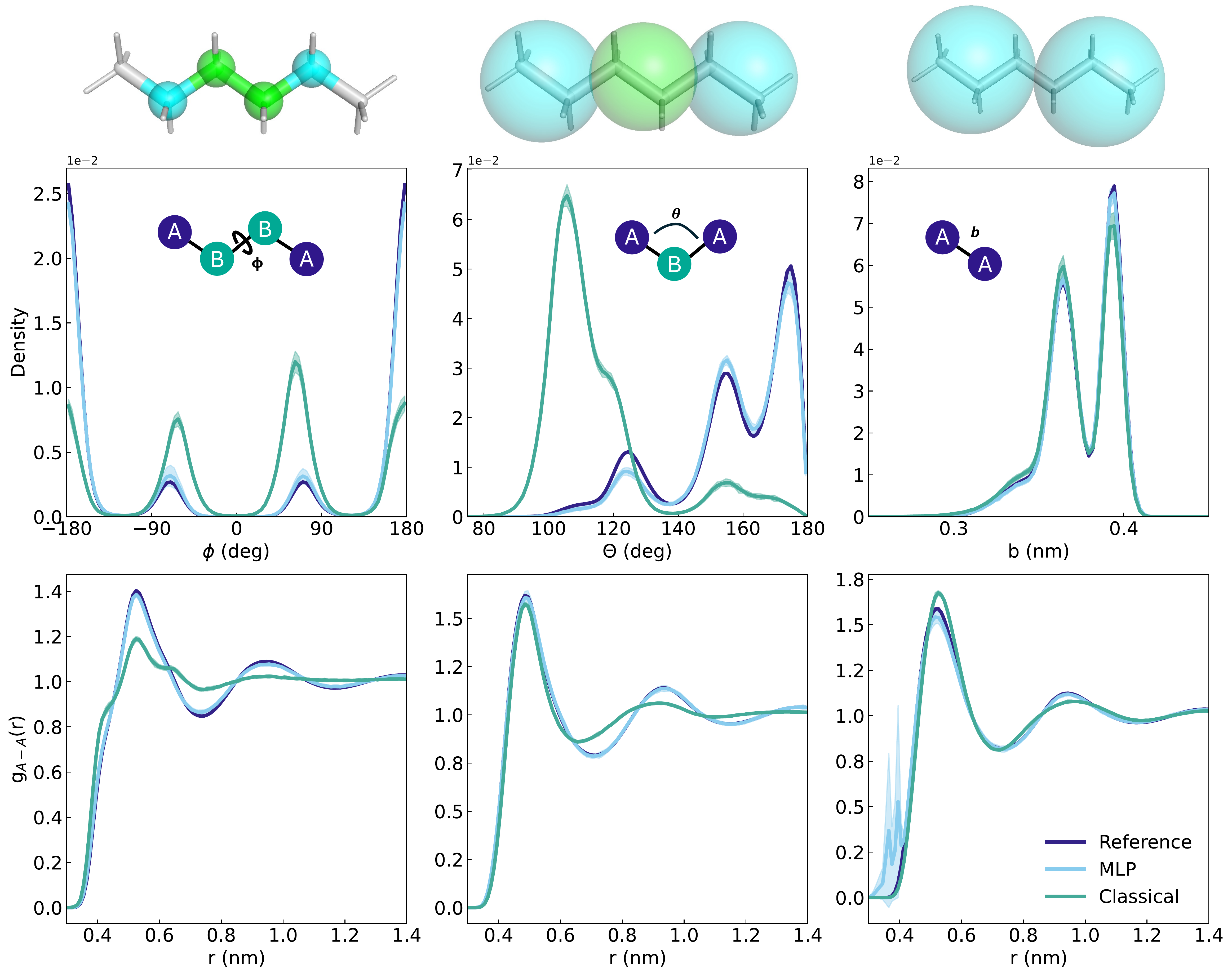}
    \caption{Structural properties of the reference, classical, MLP models for different coarse-grained representations of hexane. First row shows bonded population density metrics: Dihedral (four-site model), Angle (three-site model), Bond distance (two-site model). In case of the two-site MLP simulation, we show the nearest neighbor distance instead of bond lengths based on the initial bond list. The second row shows the RDF of A-A beads. Results show the mean $\pm$ 3 standard deviations of 10 × 1000 ps simulations.}
    \label{fig:hex_overview}
\end{figure}

We first explore liquid hexane, for which classical coarse-grained models have been extensively studied. \cite{ruhle2011hybrid, das2012multiscale, bernhardt2023stability, rudzinski2014investigation, chakraborty2020preservation} We explore three different mappings: A four-site model, where A-type beads are placed at carbons 2 and 5, and B-type beads are placed at the central carbons 3 and 4.\cite{das2012multiscale} A three-site model model, where the two consecutive carbons and their hydrogens form a bead.\cite{ruhle2011hybrid, rudzinski2014investigation, chakraborty2020preservation, bernhardt2023stability}. And finally a two-site model, in which each n-propyl end, or three consecutive carbons and associated hydrogens form a bead.\cite{das2012multiscale,chakraborty2020preservation} A visual representation can be found in Figure \ref{fig:hex_overview}. For each mapping, we derive a classical potential and a MLP using MACE.

\paragraph{Four-site Model} The results for the four-site model can be seen in the first column of Figure \ref{fig:hex_overview}. Overall, both the classical and MLP are able to replicate the shape of the dihedral population. However, the classical potential overestimates the gauche conformation of hexane with a significant increase in sampling around $\pm70$ deg. The RDF error is also much smaller for the MLP.

\paragraph{Three-site Model}
For the three-site model, the classical potential fails to sample the A-B-A angle correctly. This mismatch when using force matching with a classical potential has also been observed in other studies of hexane.\cite{rudzinski2014investigation, ruhle2009versatile} The reason why force matching fails for the classical potential is that the sum of angle, bond, and nonbonded terms does not consider out-of-plane forces that are produced by a dihedral reorientation in the atomistic model. The functional form can only capture the forces that lie in the plane in which the angle $\theta$ is defined. \cite{ruhle2009versatile}

\begin{figure}[H]
    \centering
    \includegraphics[width=16cm]{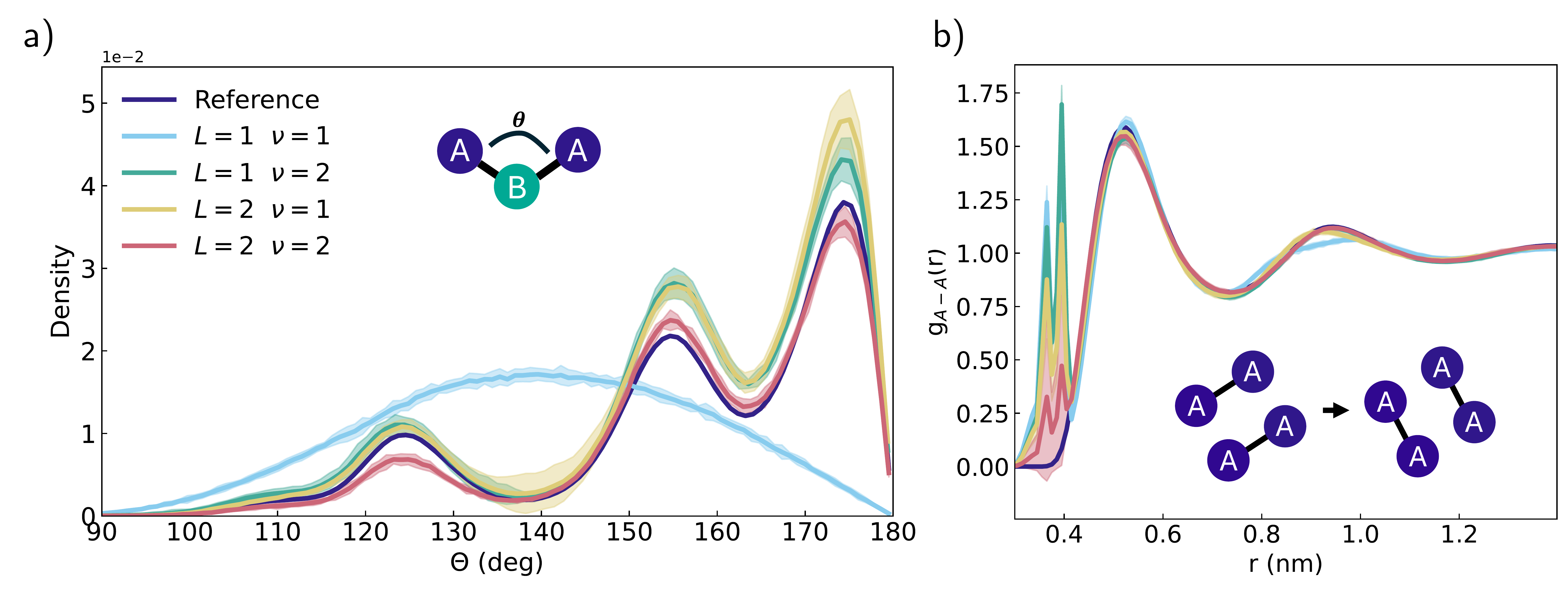}
    \caption{(a) Angular distribution of the hexane three-site model and (b) RDF of the two-site model for different correlation orders $\nu$ and number of message-passing layers $L$. In the two-site liquid hexane model, the length scales of bonded and nonbonded interactions overlap. Results show the mean $\pm$ 3 standard deviations of 10 × 1000 ps simulations.}
    \label{fig:Hexane_body_order}
\end{figure}

The MLP is able to closely match the RDF and angular distribution. We tested the effect of the number of message-passing layers $L$ and correlation order $\nu$ on the ability of the MACE model to reproduce the A-B-A angle correctly (Figure \ref{fig:Hexane_body_order}a). The total body of a MACE model is determined by $\nu$ and $L$ through $\sum_{l=0}^{L}{\nu^l}$. In the simplest case, $L=1$ and $\nu=1$, the total body order is 2 and MACE is unable to capture the angular distribution correctly. Increasing the number of layers or the correlation order allows the MACE model to capture the interaction correctly.

\paragraph{Two-site Model} 
The classical potential performs well in reproducing both bonded and nonbonded interactions of the two-site mapping. For the MLP an interesting behavior was observed: the bond partners switch over the simulation. This means, that an analysis of the bond lengths based on the initial bond list leads to a broadening and diffuse distribution. If instead the closest neighbor of each bead is used for the bond length analysis, the distribution closely matches the reference (Figure \ref{fig:hex_overview}). The impact of the bond switches is also visible in the RDF, where partners based on the initial bond list are excluded. Two clear peaks are visible at 0.35 and 0.4 nm, an artifact of the newly bonded partners. We tested the influence of $\nu$ and $L$ on this behavior but could not find any dependence (Figure \ref{fig:Hexane_body_order}b). We also tested the two-site hexane model with NequIP, observing the same result (Supporting Information).

\paragraph{Model Parameters}

\begin{table}[H]
\centering
\small
\resizebox{16cm}{!}{%
\begin{tabular}{ll|ccc|ccc}
\toprule
& $N_{\text{avg}}$ ($\propto r_{cut}$) &  & \textbf{10} &  & & \textbf{20} & \\
& $\nu$ & \textbf{1} & \textbf{2} & \textbf{3} & \textbf{1} & \textbf{2*} & \textbf{3} \\
\midrule

\multirow{5}{*}{\raisebox{-1.2\height}{\rotatebox{90}{\textbf{4-site}}}}
& Force ($\downarrow$) & 772.7 & 771.8 & 771.6 & 772.7 & 771.8 & \cellcolor{lightgreen} 771.5 \\
& Stability ($\uparrow$) & $158_{(64)}$ & $38_{(25)}$ & $851_{(272)}$ & $997_{(20)}$ & $579_{(372)}$ & \cellcolor{lightgreen} $960_{(172)}$ \\
& Speed ($\uparrow$) & \cellcolor{lightgreen} 120.2 & 113.8 & 83.3 & 81.5 & 78.1 & 62.9  \\
& $g_{A-A}$ ($\downarrow$) & -- & -- & $0.4_{(0.1)}$ & $2.8_{(1.4)}$ & $0.2_{(0.4)}$ & \cellcolor{lightgreen} $0.1_{(0.0)}$ \\
& $g_{A-B}$ ($\downarrow$) & -- & -- & $0.2_{(0.0)}$ & $5.1_{(3.1)}$ & $0.1_{(0.2)}$ & \cellcolor{lightgreen} $0.0_{(0.0)}$ \\
& $g_{B-B}$ ($\downarrow$) & -- & -- & $0.1_{(0.1)}$ & $50.3_{(31.3)}$ & $0.2_{(0.5)}$ & \cellcolor{lightgreen} $0.1_{(0.1)}$ \\
\midrule

\multirow{5}{*}{\raisebox{-1.2\height}{\rotatebox{90}{\textbf{3-site}}}}
& Force ($\downarrow$) & 418.2 & 415.6 & 414.7 & 418.2 & 415.5 & \cellcolor{lightgreen} 414.6 \\
& Stability ($\uparrow$) & \cellcolor{lightgreen} $1000_{(0)}$ & $379_{(295)}$ & $446_{(350)}$ & \cellcolor{lightgreen} $1000_{(0)}$ & $763_{(300)}$ & $419_{(331)}$ \\
& Speed ($\uparrow$) & \cellcolor{lightgreen} 151.2 & 141.8 & 107.5 & 104.6 & 99.3 & 81.2 \\
& $g_{A-A}$ ($\downarrow$) & $0.5_{(0.1)}$ & $0.2_{(0.2)}$ & $0.2_{(0.2)}$ & $0.1_{(0.2)}$ & \cellcolor{lightgreen} $0.1_{(0.3)}$ & $0.2_{(0.4)}$ \\
& $g_{A-B}$ ($\downarrow$) & $0.3_{(0.1)}$ & $0.2_{(0.2)}$ & $0.3_{(0.1)}$ & $0.1_{(0.1)}$ & \cellcolor{lightgreen} $0.1_{(0.1)}$ & $0.2_{(0.4)}$ \\
& $g_{B-B}$ ($\downarrow$) & $0.4_{(0.1)}$ & $0.6_{(0.5)}$ & $0.9_{(0.7)}$ & \cellcolor{lightgreen} $0.2_{(0.1)}$ & $0.3_{(0.4)}$ & $0.8_{(1.8)}$ \\
\midrule

\multirow{5}{*}{\raisebox{0.45\height}{\rotatebox{90}{\textbf{2-site}}}}
& Force ($\downarrow$) & 413.2 & 411.7 & 411.5 \cellcolor{lightgreen} & 413 & 411.6 & 411.5 \cellcolor{lightgreen} \\
& Stability ($\uparrow$) & \cellcolor{lightgreen} $1000_{(0)}$ & \cellcolor{lightgreen} $1000_{(0)}$ & $930_{(189)}$ & \cellcolor{lightgreen} $1000_{(0)}$ & \cellcolor{lightgreen} $1000_{(0)}$ & $790_{(271)}$ \\
& Speed ($\uparrow$) & \cellcolor{lightgreen} 179.0 & 164.0 & 134.1 & 139.8 & 131.6 & 111.8 \\
& $g_{A-A}$ ($\downarrow$) & $20.7_{(3.0)}$ & $5.2_{(3.2)}$ & $3.1_{(2.5)}$ & $20.8_{(4.2)}$ & $4.5_{(2.9)}$ & \cellcolor{lightgreen} $2.7_{(2.3)}$ \\
\bottomrule
\end{tabular}%
}
\caption{Comparison of a 2-layer MACE model with different correlation orders $\nu$ and average number of neighbors $N_{avg}$. Results are based on 50 × 1000 ps simulations. Green highlighting indicates better performance (arrows indicate direction: $\uparrow$ higher is better, $\downarrow$ lower is better). -- indicates that not enough sample statistics were available. Units: Force in meV/Å, stability in ps, $g_{A-A}$, $g_{A-B}$, $g_{B-B}$ in $10^{-3}$. Speed is listed in ns/d, based on a single 1000 ps simulation. Subscripts show standard deviations, if applicable. \textbf{*} signifies the model shown in Figure \ref{fig:hex_overview}.}
\label{tab:model_comparison}
\end{table}

Lastly, we evaluate how the cutoff radius $r_{\text{cut}}$ and the correlation order $\nu$ affect the accuracy, computational performance, and stability of the MACE potential (Table \ref{tab:model_comparison}). Overall, increasing either the correlation order or the cutoff radius reduces both RDF and force errors.\footnote{Force errors can only be compared between models using the same mapping, as the mapping itself introduces an irreducible noise \cite{wang2019machine}.} A larger cutoff radius also substantially improves stability, though at the expense of higher computational cost. As expected, less expressive models run faster. Similarly, reducing model complexity improves speed; however, classical models remain over an order of magnitude faster, achieving around 1600, 2130, and 3470 ns/d for the four-, three-, and two-site models, respectively.

\subsection{Capped Amino Acids}
\begin{figure}[H]
    \centering
    \includegraphics[width=16cm]{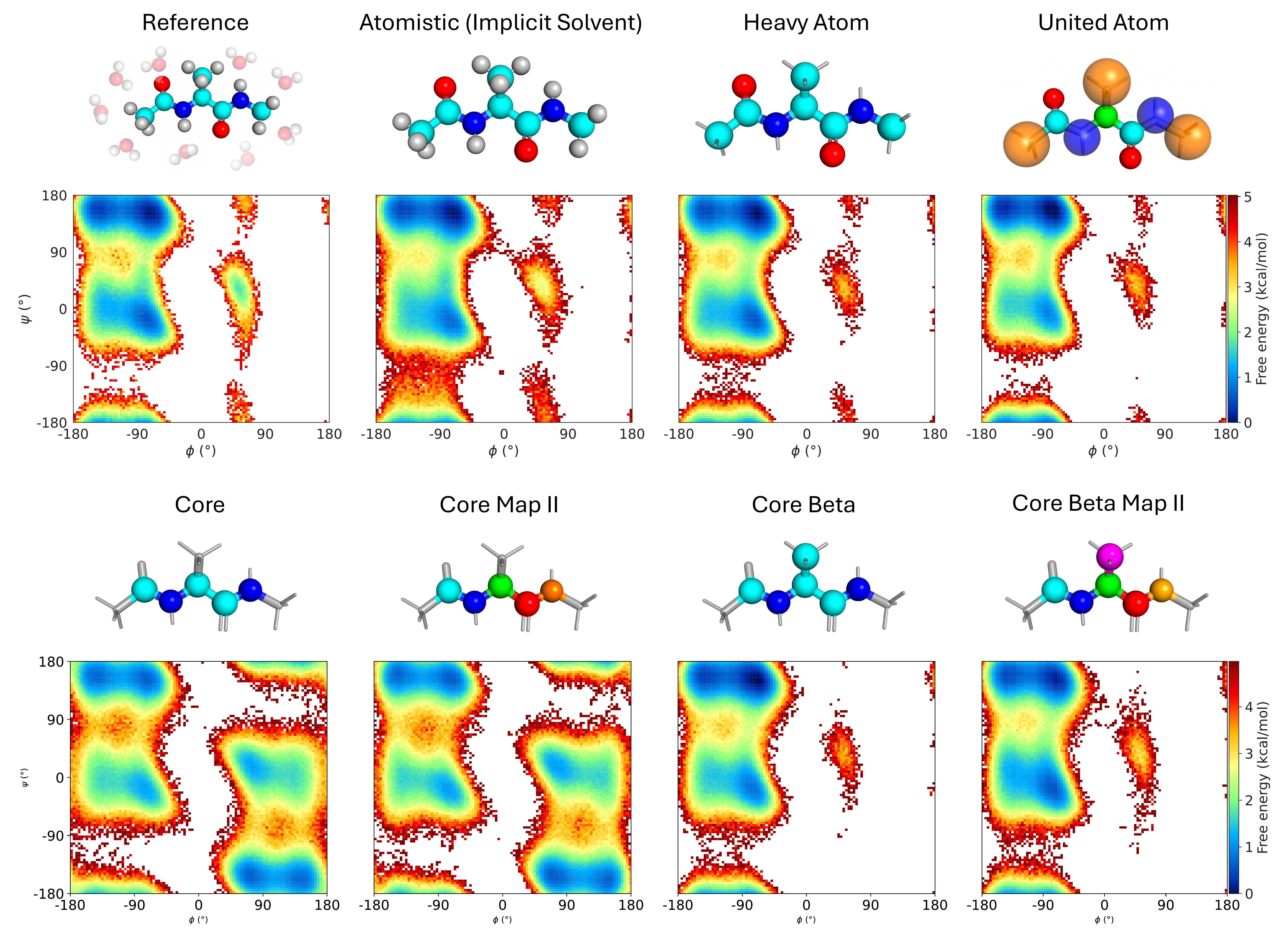}
    \caption{Results of the capped alanine CG simulations with different mappings. The top and bottom rows present the high- and low-resolution mappings, respectively. Each mapping includes a Ramachandran plot derived from 100 × 5 ns simulations.}
    \label{fig:ala2_results}
\end{figure}

In recent decades, many different coarse-grained representations of amino acids have been developed for peptide and protein-related simulations. Mappings can range from near-atomistic representations that only remove hydrogens but preserve heavy-atom movements, to single bead mappings, in which each residue is represented by one bead, for example the $C_\alpha$-atom \cite{kmiecik2016coarse}. First, we discuss the widely studied capped L-alanine system (alanine dipeptide). This system has been extensively studied using different machine learning potentials \cite{thaler2022deep, thaler2023scalable, wang2019machine, wang2020ensemble, chen2025enhanced}, as well as more recently generative models \cite{kohler2023flow, klein2025operator}. We explore nine different mappings, ranging from an atomistic implicit solvent model to five-bead mappings, which preserve only the central backbone dihedrals. An overview of the mappings can be seen in Figure \ref{fig:ala2_results}. For each mapping, we train a MACE model, and analyze the Ramachandran of the backbone dihdrals. We also perform 100 × 100 ps NVE simulations with different time steps to analyze simulation stability (Supplementary Information).

\paragraph{High-resolution Mappings}
First, we discuss the results for high-resolution mappings in which solvent molecules and potentially hydrogen atoms are removed. Retaining only the heavy atoms, or a subset of them, is a common mapping choice, as it promises to preserve atomistic dynamics and makes comparisons with atomistic models straightforward.\cite{thaler2022deep, mirarchi2024amaro, wang2019machine, wang2020ensemble, charron2025navigating} All high-resolution mappings are capable of accurately modeling the backbone dihedrals (Figure \ref{fig:ala2_results}). In the NVE simulations, the implicit solvent model remains stable up to time steps of 0.5 fs, while the United and Heavy Atom model can be run safely up to 3 fs. 

\paragraph{Low-resolution Mappings}

The Core mappings, which miss the $C_{\beta}$ carbon, show a clear point symmetry around the center (Figure \ref{fig:ala2_results} column 2). This indicates that the CG molecule during the simulation freely switches between the L- and D-enantiomer. Since both the $C_\alpha$-hydrogen and the entire side chain are removed, the formal definition of chirality is lost, and a transition between enantiomers becomes an unhindered rotation of the backbone dihedrals. Adding the $C_\beta$ atom resolves this issue, and the CG molecule stays in the original L-enantiomer. We observed no noticeable improvement in terms of structural accuracy or stability when using the more detailed species assignment of Map 2. A more subtle symmetry was observed when using a single species for all beads. In this case, the model fails to capture the direction of the molecule, which results from an indistinguishability of local environments. We provide these findings, together with evaluations on three other amino acids in the Supporting Information.

\paragraph{Chiral Inversion}
Lastly, we show that CG amino acid models can undergo chiral inversion, i.e. they transition between enantiomers, even if only a single atom is removed around the stereocenter. For that, we calculate the free energy barrier for the transition by biasing the improper $C_\alpha$-dihedral using well-tempered metadynamics. In all cases, we only train on L-alanine. We provide the exact biasing protocol in the Supporting Information.

In nature, the direct transition between enantiomers is physically blocked by a high-energy planar transition structure. Consequently, enantiomerization typically requires reactive intermediates~\cite{das2024amino, ballard2020racemisation}. The atomistic (implicit solvent) MLP accurately describes this behavior, and we were unable to observe enantiomerization up to 18 kcal/mol. We find that further increasing the bias factor renders the simulation unstable before any transition can be observed.

\begin{figure}[H]
    \centering
    \includegraphics[width=16cm]{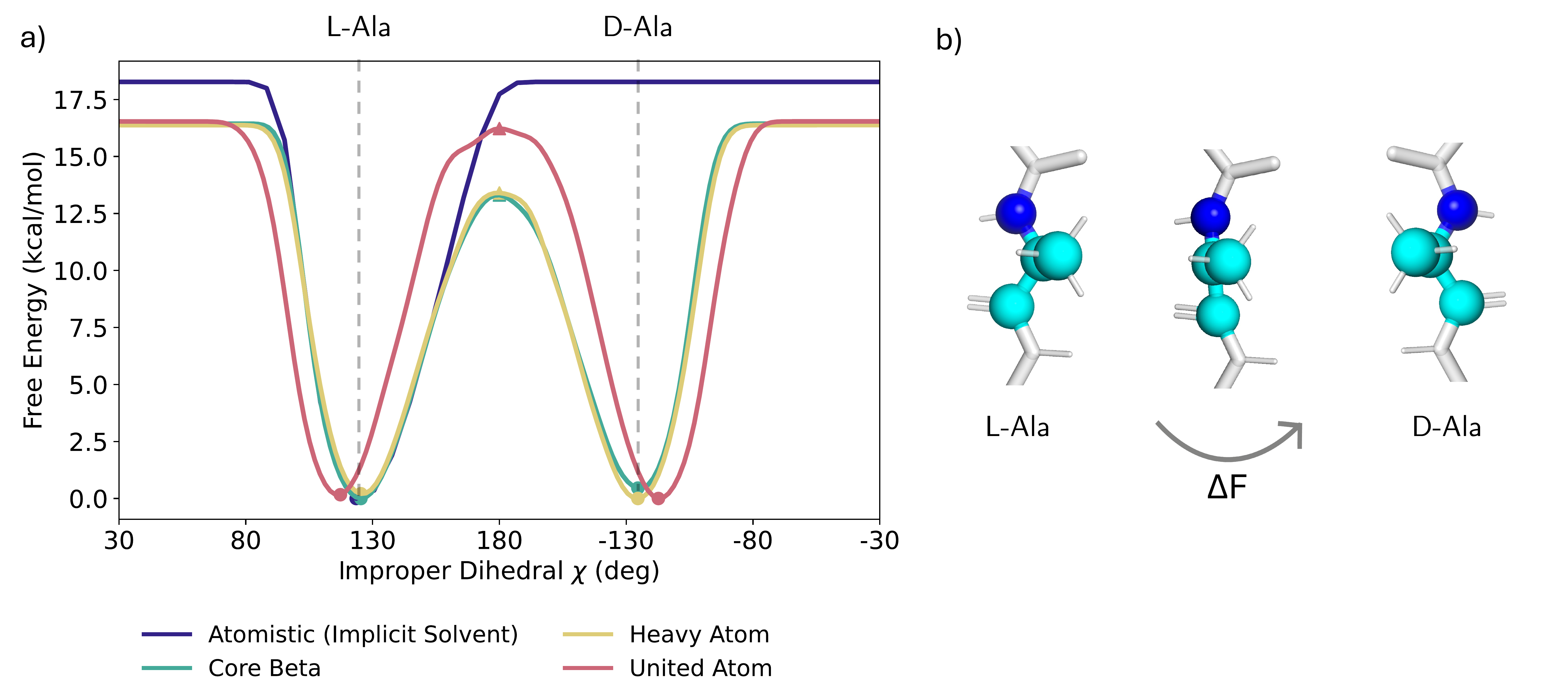}
    \caption{(a) Determined free energy barrier for enantiomerization in atomistic and coarse grained MLPs using well-tempered metadynamics along the improper $C_\alpha$-dihedral. (b) Mechanism of chiral inversion with high-energy planar transition state.}
    \label{fig:ala2_chiral_inversion}
\end{figure}

By removing one atom around the chiral center, in this case the $C_\alpha$-hydrogen, enantiomerization becomes accessible (Figure \ref{fig:ala2_chiral_inversion}a). For the Heavy Atom and Core Beta mappings, we find a transition barrier of approximately 13 kcal/mol. The United Atom mapping presents a slightly larger barrier of 16 kcal/mol. This increase arises because merging heavy atoms with their hydrogens shifts the bead COM away from the stereocenter, thereby stabilizing the initial configuration. This also slightly shifts the minima of the improper dihedral.

The calculated free energy barriers are very high compared to the thermal energy available of $\sim$ 0.6 kcal/mol at 300 K, making transitions very unlikely. At this barrier height, transitions take on the order of micro- to milliseconds. However, increasing the temperature would drastically increase the transition rate, further restricting the state point dependence of the CG potential.

\subsection{Helix Formation of Polyalanine}

\begin{figure}[H]
    \centering
    \includegraphics[width=16cm]{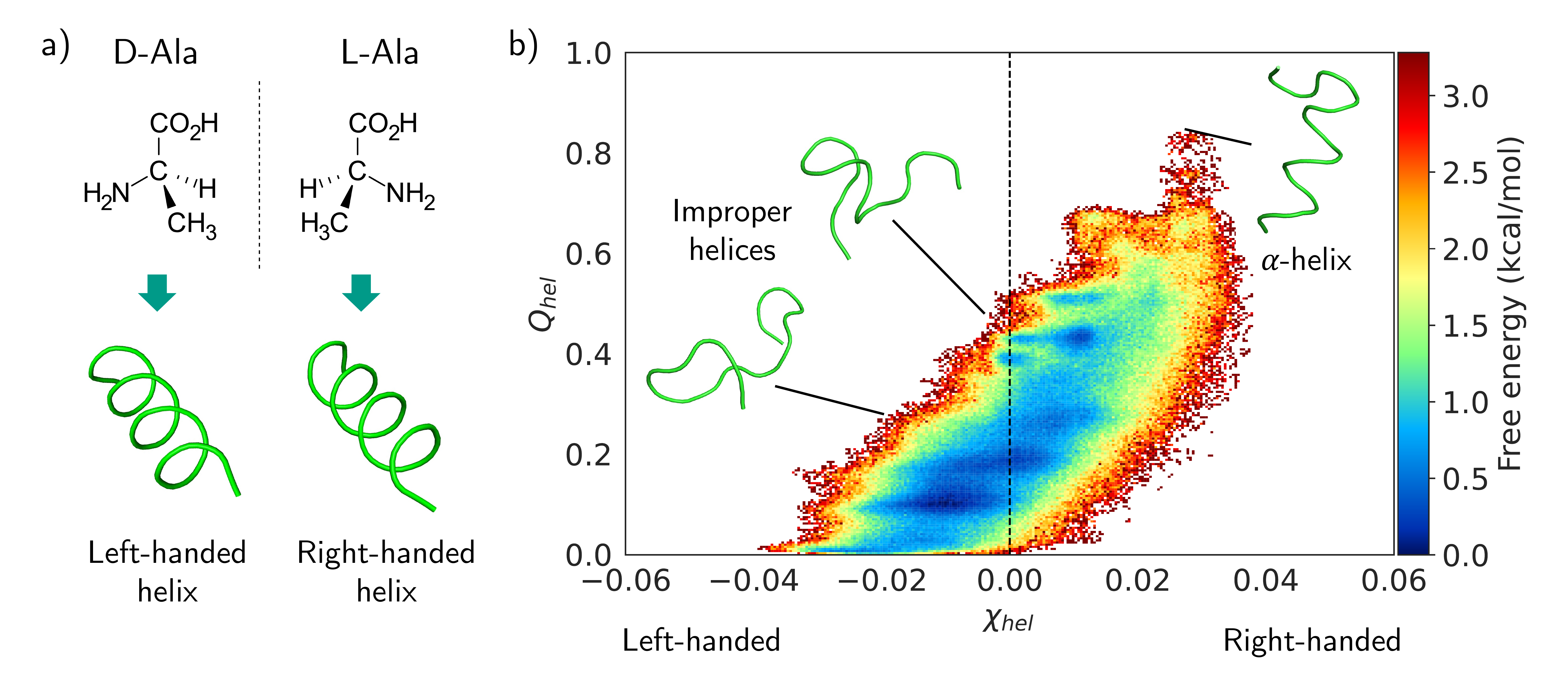}
    \caption{(a) A chain of D-Alanine forms left-handed helices, while the naturally occurring L-Alanine forms right-handed helices. (b) Helicity and handedness of a 500 ns reference simulation of a capped 15-mer L-alanine.}
    \label{fig:ala15_ref_overview}
\end{figure}

As a last experiment, we extend the low resolution mappings from capped alanine to a capped 15-mer alanine. Polyalanine systems are good examples of peptides that spontaneously form alpha helices \cite{hazel2014thermodynamics, rudzinski2015bottom, kuczera2021length, zhou2007coarse, chakrabarti2001interrelationships}. While naturally occurring L-amino acid peptides form right-handed helices (Figure \ref{fig:ala15_ref_overview}a)\cite{sidorova2021protein}, it has been experimentally shown that a D-alanine peptide forms the mirror image, a left-handed helix.\cite{milton1992} 

We tested three different mappings: the Core and Core Beta mapping from the capped alanine example, and a $C_\alpha$ mapping, which only preserves the $C_\alpha$ atom of each alanine residue. The $C_\alpha$ mapping is a popular choice for modeling large protein systems.\cite{tozzini2006mapping, yap2008coarse, kmiecik2016coarse, mioduszewski2023contact} In the $C_\alpha$ mapping, we gave each $C_\alpha$ a different species. We also performed experiments on three other possible choices: a single species, alternating species, and a symmetric species arrangement. Using unique or alternating species causes unphysical bond switches and ultimately instabilities. We all the results of all $C_\alpha$ mappings in the Supporting Information.

\begin{figure}[H]
    \centering
    \includegraphics[width=16cm]{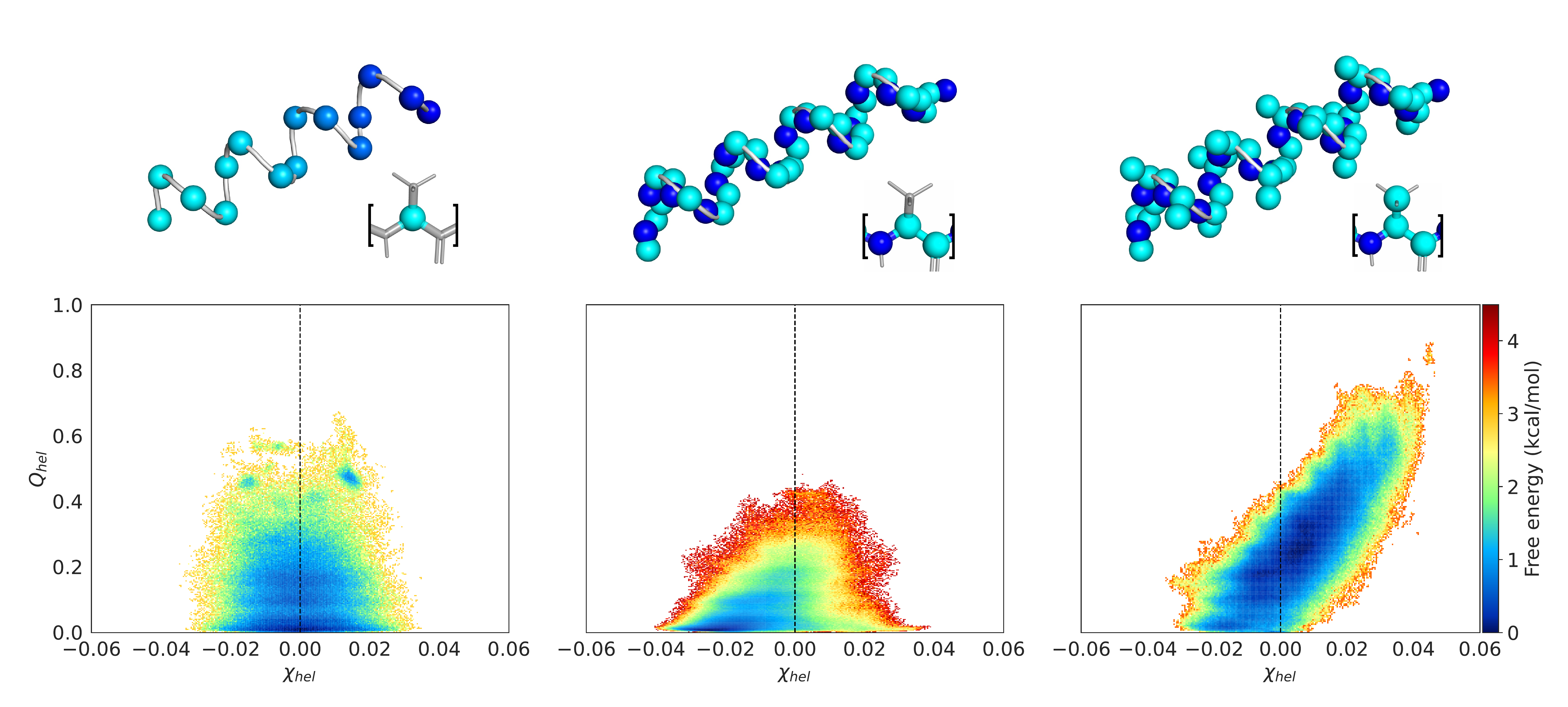}
    \caption{Helicity and handedness of 100 × 5 ns trajectories based on different CG mappings. Left: $C_\alpha$, where each bead has a unique species embedding. Middle: Core. Right: Core Beta. The top row shows a graphical representation of the ideal helix and per residue mapping.}
    \label{fig:ala15_MACE_overview}
\end{figure}

The helix formation of polyalanine can be described via two order parameters: an index of alpha-helicity $Q_{hel}$ (Equation \ref{helicity_eq}) and a chirality/handedness index $\chi_{hel}$ (Equation \ref{handedness_eq}). In the reference atomistic simulation, a clear population of partially folded right-handed alpha helices can be seen ($\chi_{hel}>0$ and $Q_{hel}>0$). Although left-handed helices are visible, these are not proper alpha-helices (Figure \ref{fig:ala15_ref_overview}b).

When investigating the helicity and handedness of the CG simulations, clear symmetries can be observed (Figure \ref{fig:ala15_MACE_overview}). The $C_\alpha$ mapping is completely symmetric to $\chi_{hel}=0$, which means that the model does not have a preference for helix handedness. The Core mapping also shows increased sampling of incorrect left-handed helices, even though a preference for the true right-handed helices can be observed. This might be because the starting frames were taken from the L-Polylanine reference simulation and are thus favorably oriented to form the right-handed helix. The Core Beta mapping captures the correct helix formation.

\section{Conclusions}

Our systematic evaluation of CG models for liquid hexane, amino acids, and polyalanine reveals that the interplay between mapping and model architecture greatly impacts the learned representation of equivariant MLPs. Classical potentials rely on fixed functional forms, which work well for the two-site hexane model but lack the flexibility to capture the more complex many-body correlations present in the three- and four-site liquid hexane models. In contrast, MLPs like MACE overcome this barrier by learning many-body interactions directly from data. MACE performs well for the three- and four-site mappings of hexane, accurately recovering both bonded and nonbonded interactions. 

Because MLPs do not include explicit topological information, they cannot distinguish bonded neighbors from nonbonded neighbors. In atomistic systems, these interactions are naturally separable because of their different length scales.\cite{loose2023coarse} In contrast, for CG representations such as the two-site liquid hexane or polyalanine $C_\alpha$ model, we showed that these length scales overlap. In these cases, the model fails to distinguish particles, which leads to unphysical bond permutations and ultimately instabilities, in case of the polyalanine $C_\alpha$ model.

We showed that high-resolution mappings for amino acids can accurately capture atomistic backbone dynamics. However, when further coarsening the representation, we observed spurious symmetries. While equivariant MLPs can distinguish between enantiomers through parity-sensitive features, they will assign the same energy to both mirror images if the output is constrained to a scalar energy. In atomistic systems, this is a correct and desired symmetry, since the transition between enantiomers is correctly modeled with a practically infinite energy barrier. However, we showed that upon removal of the $C_\alpha$-hydrogen, enantiomerization becomes accessible through a high-energy transition state. If additionally the side chain is removed, the molecule loses its formal chirality and a symmetric FES is obtained. Additionally, bead species have to be chosen carefully, as indistinguishability of local environments can lead to further symmetries.

We confirmed key findings with NequIP, a different E(3)-equivariant MLP.\cite{batzner20223} Our findings likely also extend to invariant graph neural networks, such as SchNet\cite{schutt2018schnet} or DimeNet\cite{gasteiger2020directional}, which rely on up to three-body scalar invariants (distances and angles). These models are blind to chirality, making them susceptible to the enantiomerization we observed.\cite{dumitrescu20243, adams2021learning} Finally, bond permutations are likely to affect any architecture that constructs neighborhoods purely based on geometric information, without topological information.

Overall, equivariant MLPs outperform classical potentials regarding expressivity and data efficiency, theoretically allowing them to learn the potential of mean force for any mapping. In practice, however, our results show that the applicability is limited by mapping-induced artifacts. In scenarios where topological conservation is secondary, such as the liquid hexane systems, MLPs excel at reproducing structural distributions while requiring minimal preparation. However, caution is required when applying MLPs to systems where topology is critical, such as for CG protein models. In these contexts, current MLPs are generally restricted to implicit solvent or Heavy Atom models. Coarser representations, such as the Core and Core Beta mapping require thought about possible symmetries and overlapping length scales. An even coarser representation, for example the protein $C_\alpha$-model, risks introducing instabilities, overlapping length scales, or spurious symmetries. Such coarse representations are unsuitable for  MLPs in their current state, and will necessitate some form of topology information or regularization via prior energy terms.

\begin{acknowledgement}
Funded by the European Union. Views and opinions expressed are however those of the
author(s) only and do not necessarily reflect those of the European Union or the European Research Council
Executive Agency. Neither the European Union nor the granting authority can be held responsible for them. This
work was funded by the ERC (StG SupraModel) - 101077842.
\end{acknowledgement}


\section{Data and Software Availability}
The code and data supporting this study are publicly available at \url{https://github.com/tummfm/CG-Mapping-Benchmark}. The training framework \texttt{chemtrain} is publicly available at \url{https://github.com/tummfm/chemtrain}. 

\bibliography{ref}

@article{robertson2015improved,
  title={Improved peptide and protein torsional energetics with the OPLS-AA force field},
  author={Robertson, Michael J and Tirado-Rives, Julian and Jorgensen, William L},
  journal={Journal of chemical theory and computation},
  volume={11},
  number={7},
  pages={3499--3509},
  year={2015},
  publisher={ACS Publications}
}

@article{lindorff2010improved,
  title={Improved side-chain torsion potentials for the Amber ff99SB protein force field},
  author={Lindorff-Larsen, Kresten and Piana, Stefano and Palmo, Kim and Maragakis, Paul and Klepeis, John L and Dror, Ron O and Shaw, David E},
  journal={Proteins: Structure, Function, and Bioinformatics},
  volume={78},
  number={8},
  pages={1950--1958},
  year={2010},
  publisher={Wiley Online Library}
}

@article{ruhle2011hybrid,
  title={Hybrid approaches to coarse-graining using the VOTCA package: liquid hexane},
  author={R{\"u}hle, Victor and Junghans, Christoph},
  journal={Macromolecular Theory and Simulations},
  volume={20},
  number={7},
  pages={472--477},
  year={2011},
  publisher={Wiley Online Library}
}

@article{ruhle2009versatile,
  title={Versatile object-oriented toolkit for coarse-graining applicationsf},
  author={Ruhle, Victor and Junghans, Christoph and Lukyanov, Alexander and Kremer, Kurt and Andrienko, Denis},
  journal={Journal of chemical theory and computation},
  volume={5},
  number={12},
  pages={3211--3223},
  year={2009},
  publisher={ACS Publications}
}

@article{wang2019machine,
  title={Machine learning of coarse-grained molecular dynamics force fields},
  author={Wang, Jiang and Olsson, Simon and Wehmeyer, Christoph and P{\'e}rez, Adri{\`a} and Charron, Nicholas E and De Fabritiis, Gianni and No{\'e}, Frank and Clementi, Cecilia},
  journal={ACS central science},
  volume={5},
  number={5},
  pages={755--767},
  year={2019},
  publisher={ACS Publications}
}

@article{thaler2023scalable,
  title={Scalable Bayesian uncertainty quantification for neural network potentials: promise and pitfalls},
  author={Thaler, Stephan and Doehner, Gregor and Zavadlav, Julija},
  journal={Journal of Chemical Theory and Computation},
  volume={19},
  number={14},
  pages={4520--4532},
  year={2023},
  publisher={ACS Publications}
}

@article{charron2025navigating,
  title={Navigating protein landscapes with a machine-learned transferable coarse-grained model},
  author={Charron, Nicholas E and Bonneau, Klara and Pasos-Trejo, Aldo S and Guljas, Andrea and Chen, Yaoyi and Musil, F{\'e}lix and Venturin, Jacopo and Gusew, Daria and Zaporozhets, Iryna and Kr{\"a}mer, Andreas and others},
  journal={Nature chemistry},
  pages={1--9},
  year={2025},
  publisher={Nature Publishing Group UK London}
}

@article{thaler2022deep,
  title={Deep coarse-grained potentials via relative entropy minimization},
  author={Thaler, Stephan and Stupp, Maximilian and Zavadlav, Julija},
  journal={The Journal of Chemical Physics},
  volume={157},
  number={24},
  year={2022},
  publisher={AIP Publishing}
}

@article{kohler2023flow,
  title={Flow-matching: Efficient coarse-graining of molecular dynamics without forces},
  author={Kohler, Jonas and Chen, Yaoyi and Kramer, Andreas and Clementi, Cecilia and No{\'e}, Frank},
  journal={Journal of Chemical Theory and Computation},
  volume={19},
  number={3},
  pages={942--952},
  year={2023},
  publisher={ACS Publications}
}

@article{wang2020ensemble,
  title={Ensemble learning of coarse-grained molecular dynamics force fields with a kernel approach},
  author={Wang, Jiang and Chmiela, Stefan and M{\"u}ller, Klaus-Robert and No{\'e}, Frank and Clementi, Cecilia},
  journal={The Journal of chemical physics},
  volume={152},
  number={19},
  year={2020},
  publisher={AIP Publishing}
}

@article{klein2025operator,
  title={Operator forces for coarse-grained molecular dynamics},
  author={Klein, Leon and Kelkar, Atharva and Durumeric, Aleksander and Chen, Yaoyi and Clementi, Cecilia and No{\'e}, Frank},
  journal={The Journal of Chemical Physics},
  volume={163},
  number={10},
  year={2025},
  publisher={AIP Publishing}
}

@article{chakrabarti2001interrelationships,
  title={The interrelationships of side-chain and main-chain conformations in proteins},
  author={Chakrabarti, Pinak and Pal, Debnath},
  journal={Progress in biophysics and molecular biology},
  volume={76},
  number={1-2},
  pages={1--102},
  year={2001},
  publisher={Elsevier}
}

@article{izvekov2005multiscale,
  title={A multiscale coarse-graining method for biomolecular systems},
  author={Izvekov, Sergei and Voth, Gregory A},
  journal={The Journal of Physical Chemistry B},
  volume={109},
  number={7},
  pages={2469--2473},
  year={2005},
  publisher={ACS Publications}
}

@article{souza2021martini,
  title={Martini 3: a general purpose force field for coarse-grained molecular dynamics},
  author={Souza, Paulo CT and Alessandri, Riccardo and Barnoud, Jonathan and Thallmair, Sebastian and Faustino, Ignacio and Gr{\"u}newald, Fabian and Patmanidis, Ilias and Abdizadeh, Haleh and Bruininks, Bart MH and Wassenaar, Tsjerk A and others},
  journal={Nature methods},
  volume={18},
  number={4},
  pages={382--388},
  year={2021},
  publisher={Nature Publishing Group US New York}
}

@article{das2012multiscale,
  title={The multiscale coarse-graining method. X. Improved algorithms for constructing coarse-grained potentials for molecular systems},
  author={Das, Avisek and Lu, Lanyuan and Andersen, Hans C and Voth, Gregory A},
  journal={The Journal of Chemical Physics},
  volume={136},
  number={19},
  year={2012},
  publisher={AIP Publishing}
}

@article{chakraborty2020preservation,
  title={Is preservation of symmetry necessary for coarse-graining?},
  author={Chakraborty, Maghesree and Xu, Jinyu and White, Andrew D},
  journal={Physical Chemistry Chemical Physics},
  volume={22},
  number={26},
  pages={14998--15005},
  year={2020},
  publisher={Royal Society of Chemistry}
}

@article{bernhardt2023stability,
  title={Stability, speed, and constraints for structural coarse-graining in VOTCA},
  author={Bernhardt, Marvin P and Hanke, Martin and van der Vegt, Nico FA},
  journal={Journal of Chemical Theory and Computation},
  volume={19},
  number={2},
  pages={580--595},
  year={2023},
  publisher={ACS Publications}
}

@article{kuczera2021length,
  title={Length dependent folding kinetics of alanine-based helical peptides from optimal dimensionality reduction},
  author={Kuczera, Krzysztof and Szoszkiewicz, Robert and He, Jinyan and Jas, Gouri S},
  journal={Life},
  volume={11},
  number={5},
  pages={385},
  year={2021},
  publisher={MDPI}
}

@article{fu2022forces,
  title={Forces are not enough: Benchmark and critical evaluation for machine learning force fields with molecular simulations},
  author={Fu, Xiang and Wu, Zhenghao and Wang, Wujie and Xie, Tian and Keten, Sinan and Gomez-Bombarelli, Rafael and Jaakkola, Tommi},
  journal={arXiv preprint arXiv:2210.07237},
  year={2022}
}

@article{rudzinski2014investigation,
  title={Investigation of coarse-grained mappings via an iterative generalized Yvon--Born--Green method},
  author={Rudzinski, Joseph F and Noid, William G},
  journal={The Journal of Physical Chemistry B},
  volume={118},
  number={28},
  pages={8295--8312},
  year={2014},
  publisher={ACS Publications}
}

@article{mirarchi2024amaro,
  title={AMARO: All heavy-atom transferable neural network potentials of protein thermodynamics},
  author={Mirarchi, Antonio and Pel{\'a}ez, Ra{\'u}l P and Simeon, Guillem and De Fabritiis, Gianni},
  journal={Journal of Chemical Theory and Computation},
  volume={20},
  number={22},
  pages={9871--9878},
  year={2024},
  publisher={ACS Publications}
}

@article{wang2019coarse,
  title={Coarse-graining auto-encoders for molecular dynamics},
  author={Wang, Wujie and G{\'o}mez-Bombarelli, Rafael},
  journal={npj Computational Materials},
  volume={5},
  number={1},
  pages={125},
  year={2019},
  publisher={Nature Publishing Group UK London}
}

@article{noid2008multiscale,
  title={The multiscale coarse-graining method. I. A rigorous bridge between atomistic and coarse-grained models},
  author={Noid, William George and Chu, Jhih-Wei and Ayton, Gary S and Krishna, Vinod and Izvekov, Sergei and Voth, Gregory A and Das, Avisek and Andersen, Hans C},
  journal={The Journal of chemical physics},
  volume={128},
  number={24},
  year={2008},
  publisher={AIP Publishing}
}

@article{rudzinski2015bottom,
  title={Bottom-up coarse-graining of peptide ensembles and helix--coil transitions},
  author={Rudzinski, Joseph F and Noid, William G},
  journal={Journal of chemical theory and computation},
  volume={11},
  number={3},
  pages={1278--1291},
  year={2015},
  publisher={ACS Publications}
}

@article{gromacs,
  title={GROMACS: High performance molecular simulations through multi-level parallelism from laptops to supercomputers},
  author={Abraham, Mark James and Murtola, Teemu and Schulz, Roland and P{\'a}ll, Szil{\'a}rd and Smith, Jeremy C and Hess, Berk and Lindahl, Erik},
  journal={SoftwareX},
  volume={1},
  pages={19--25},
  year={2015},
  publisher={Elsevier}
}

@inproceedings{MACE,
 author = {Batatia, Ilyes and Kovacs, David P and Simm, Gregor and Ortner, Christoph and Csanyi, Gabor},
 booktitle = {Advances in Neural Information Processing Systems},
 editor = {S. Koyejo and S. Mohamed and A. Agarwal and D. Belgrave and K. Cho and A. Oh},
 pages = {11423--11436},
 publisher = {Curran Associates, Inc.},
 title = {MACE: Higher Order Equivariant Message Passing Neural Networks for Fast and Accurate Force Fields},
 url = {https://proceedings.neurips.cc/paper_files/paper/2022/file/4a36c3c51af11ed9f34615b81edb5bbc-Paper-Conference.pdf},
 volume = {35},
 year = {2022}
}

@BOOK{Voth2008-zb,
  title     = "Coarse-graining of condensed phase and biomolecular systems",
  editor    = "Voth, Gregory A",
  publisher = "CRC Press",
  month     =  sep,
  year      =  2008,
  address   = "Boca Raton, FL"
}

@article{fuchs2025chemtrain,
  title={chemtrain: Learning deep potential models via automatic differentiation and statistical physics},
  author={Fuchs, Paul and Thaler, Stephan and R{\"o}cken, Sebastien and Zavadlav, Julija},
  journal={Computer Physics Communications},
  volume={310},
  pages={109512},
  year={2025},
  publisher={Elsevier}
}

@article{fuchs2025chemtraindeploy,
  title={chemtrain-deploy: A parallel and scalable framework for machine learning potentials in million-atom MD simulations},
  author={Fuchs, Paul and Chen, Weilong and Thaler, Stephan and Zavadlav, Julija},
  journal={Journal of Chemical Theory and Computation},
  volume={21},
  number={15},
  pages={7550--7560},
  year={2025},
  publisher={ACS Publications}
}

@article{john2017many,
  title={Many-body coarse-grained interactions using Gaussian approximation potentials},
  author={John, ST and Cs{\'a}nyi, G{\'a}bor},
  journal={The Journal of Physical Chemistry B},
  volume={121},
  number={48},
  pages={10934--10949},
  year={2017},
  publisher={ACS Publications}
}

@article{jin2022bottom,
  title={Bottom-up coarse-graining: Principles and perspectives},
  author={Jin, Jaehyeok and Pak, Alexander J and Durumeric, Aleksander EP and Loose, Timothy D and Voth, Gregory A},
  journal={Journal of chemical theory and computation},
  volume={18},
  number={10},
  pages={5759--5791},
  year={2022},
  publisher={ACS Publications}
}

@article{noid2024rigorous,
  title={Rigorous progress in coarse-graining},
  author={Noid, WG and Szukalo, Ryan J and Kidder, Katherine M and Lesniewski, Maria C},
  journal={Annual review of physical chemistry},
  volume={75},
  number={1},
  pages={21--45},
  year={2024},
  publisher={Annual Reviews}
}

@article{babadi2006coarse,
  title={Coarse-grained interaction potentials for anisotropic molecules},
  author={Babadi, M and Everaers, R and Ejtehadi, MR},
  journal={The Journal of chemical physics},
  volume={124},
  number={17},
  year={2006},
  publisher={AIP Publishing}
}

@article{gay1981modification,
  title={Modification of the overlap potential to mimic a linear site-site potential},
  author={Gay, JG and Berne, Bruce J},
  journal={Journal of chemical physics},
  volume={74},
  number={6},
  pages={3316--3319},
  year={1981}
}

@article{noid2013perspective,
  title={Perspective: Coarse-grained models for biomolecular systems},
  author={Noid, William George},
  journal={The Journal of chemical physics},
  volume={139},
  number={9},
  year={2013},
  publisher={AIP Publishing}
}

@article{noid2023perspective,
  title={Perspective: Advances, challenges, and insight for predictive coarse-grained models},
  author={Noid, William George},
  journal={The Journal of Physical Chemistry B},
  volume={127},
  number={19},
  pages={4174--4207},
  year={2023},
  publisher={ACS Publications}
}

@article{kmiecik2016coarse,
  title={Coarse-grained protein models and their applications},
  author={Kmiecik, Sebastian and Gront, Dominik and Kolinski, Michal and Wieteska, Lukasz and Dawid, Aleksandra Elzbieta and Kolinski, Andrzej},
  journal={Chemical reviews},
  volume={116},
  number={14},
  pages={7898--7936},
  year={2016},
  publisher={ACS Publications}
}

@article{thaler2021learning,
  title={Learning neural network potentials from experimental data via Differentiable Trajectory Reweighting},
  author={Thaler, Stephan and Zavadlav, Julija},
  journal={Nature communications},
  volume={12},
  number={1},
  pages={6884},
  year={2021},
  publisher={Nature Publishing Group UK London}
}

@article{sidorova2021protein,
  title={Protein helical structures: Defining handedness and localization features},
  author={Sidorova, Alla E and Malyshko, Ekaterina V and Lutsenko, Aleksey O and Shpigun, Denis K and Bagrova, Olga E},
  journal={Symmetry},
  volume={13},
  number={5},
  pages={879},
  year={2021},
  publisher={MDPI}
}

@article{batzner20223,
  title={E (3)-equivariant graph neural networks for data-efficient and accurate interatomic potentials},
  author={Batzner, Simon and Musaelian, Albert and Sun, Lixin and Geiger, Mario and Mailoa, Jonathan P and Kornbluth, Mordechai and Molinari, Nicola and Smidt, Tess E and Kozinsky, Boris},
  journal={Nature communications},
  volume={13},
  number={1},
  pages={2453},
  year={2022},
  publisher={Nature Publishing Group UK London}
}

@article{musaelian2023learning,
  title={Learning local equivariant representations for large-scale atomistic dynamics},
  author={Musaelian, Albert and Batzner, Simon and Johansson, Anders and Sun, Lixin and Owen, Cameron J and Kornbluth, Mordechai and Kozinsky, Boris},
  journal={Nature Communications},
  volume={14},
  number={1},
  pages={579},
  year={2023},
  publisher={Nature Publishing Group UK London}
}

@article{kocer2022neural,
  title={Neural network potentials: A concise overview of methods},
  author={Kocer, Emir and Ko, Tsz Wai and Behler, J{\"o}rg},
  journal={Annual review of physical chemistry},
  volume={73},
  number={1},
  pages={163--186},
  year={2022},
  publisher={Annual Reviews}
}

@article{thiemann2024introduction,
  title={Introduction to machine learning potentials for atomistic simulations},
  author={Thiemann, Fabian L and O’neill, Niamh and Kapil, Venkat and Michaelides, Angelos and Schran, Christoph},
  journal={Journal of Physics: Condensed Matter},
  volume={37},
  number={7},
  pages={073002},
  year={2024},
  publisher={IOP Publishing}
}

@article{durumeric2023machine,
  title={Machine learned coarse-grained protein force-fields: Are we there yet?},
  author={Durumeric, Aleksander EP and Charron, Nicholas E and Templeton, Clark and Musil, F{\'e}lix and Bonneau, Klara and Pasos-Trejo, Aldo S and Chen, Yaoyi and Kelkar, Atharva and No{\'e}, Frank and Clementi, Cecilia},
  journal={Current opinion in structural biology},
  volume={79},
  pages={102533},
  year={2023},
  publisher={Elsevier}
}

@article{ge2025coarse,
  title={Coarse-grained models for ionic liquids and applications to biological and electrochemical systems},
  author={Ge, Yang and Zhu, Qiang and Wang, Xueping and Ma, Jing},
  journal={Industrial Chemistry \& Materials},
  year={2025},
  publisher={Royal Society of Chemistry}
}

@article{drautz2019atomic,
  title={Atomic cluster expansion for accurate and transferable interatomic potentials},
  author={Drautz, Ralf},
  journal={Physical Review B},
  volume={99},
  number={1},
  pages={014104},
  year={2019},
  publisher={APS}
}

@article{loose2023coarse,
  title={Coarse-graining with equivariant neural networks: A path toward accurate and data-efficient models},
  author={Loose, Timothy D and Sahrmann, Patrick G and Qu, Thomas S and Voth, Gregory A},
  journal={The Journal of Physical Chemistry B},
  volume={127},
  number={49},
  pages={10564--10572},
  year={2023},
  publisher={ACS Publications}
}

@article{
milton1992,
author = {R. C. deL. Milton  and S. C. F. Milton  and S. B. H. Kent },
title = {Total Chemical Synthesis of a D-Enzyme: The Enantiomers of HIV-1 Protease Show Reciprocal Chiral Substrate Specificity},
journal = {Science},
volume = {256},
number = {5062},
pages = {1445-1448},
year = {1992},
doi = {10.1126/science.1604320},
URL = {https://www.science.org/doi/abs/10.1126/science.1604320},
eprint = {https://www.science.org/doi/pdf/10.1126/science.1604320}}

@article{zhou2007coarse,
  title={Coarse-grained peptide modeling using a systematic multiscale approach},
  author={Zhou, Jian and Thorpe, Ian F and Izvekov, Sergey and Voth, Gregory A},
  journal={Biophysical journal},
  volume={92},
  number={12},
  pages={4289--4303},
  year={2007},
  publisher={Elsevier}
}

@article{hazel2014thermodynamics,
  title={Thermodynamics of deca-alanine folding in water},
  author={Hazel, Anthony and Chipot, Christophe and Gumbart, James C},
  journal={Journal of chemical theory and computation},
  volume={10},
  number={7},
  pages={2836--2844},
  year={2014},
  publisher={ACS Publications}
}

@article{freud2020,
    title = {freud: A Software Suite for High Throughput
             Analysis of Particle Simulation Data},
    author = {Vyas Ramasubramani and
              Bradley D. Dice and
              Eric S. Harper and
              Matthew P. Spellings and
              Joshua A. Anderson and
              Sharon C. Glotzer},
    journal = {Computer Physics Communications},
    volume = {254},
    pages = {107275},
    year = {2020},
    issn = {0010-4655},
    doi = {https://doi.org/10.1016/j.cpc.2020.107275},
    url = {http://www.sciencedirect.com/science/article/pii/S0010465520300916},
}

@article{thompson2022lammps,
  title={LAMMPS-a flexible simulation tool for particle-based materials modeling at the atomic, meso, and continuum scales},
  author={Thompson, Aidan P and Aktulga, H Metin and Berger, Richard and Bolintineanu, Dan S and Brown, W Michael and Crozier, Paul S and In't Veld, Pieter J and Kohlmeyer, Axel and Moore, Stan G and Nguyen, Trung Dac and others},
  journal={Computer physics communications},
  volume={271},
  pages={108171},
  year={2022},
  publisher={Elsevier}
}

@article{dumitrescu20243,
  title={E (3)-equivariant models cannot learn chirality: Field-based molecular generation},
  author={Dumitrescu, Alexandru and Korpela, Dani and Heinonen, Markus and Verma, Yogesh and Iakovlev, Valerii and Garg, Vikas and L{\"a}hdesm{\"a}ki, Harri},
  journal={arXiv preprint arXiv:2402.15864},
  year={2024}
}

@article{behler2021machine,
  title={Machine learning potentials for extended systems: a perspective},
  author={Behler, J{\"o}rg and Cs{\'a}nyi, G{\'a}bor},
  journal={The European Physical Journal B},
  volume={94},
  number={7},
  pages={142},
  year={2021},
  publisher={Springer}
}

@article{yang2023slicing,
  title={Slicing and dicing: Optimal coarse-grained representation to preserve molecular kinetics},
  author={Yang, Wangfei and Templeton, Clark and Rosenberger, David and Bittracher, Andreas and N{"u}ske, Feliks and No{\'e}, Frank and Clementi, Cecilia},
  journal={ACS Central Science},
  volume={9},
  number={2},
  pages={186--196},
  year={2023},
  publisher={ACS Publications}
}

@article{chen2025enhanced,
  title={Enhanced Sampling for Efficient Learning of Coarse-Grained Machine Learning Potentials},
  author={Chen, Weilong and G{\"o}rlich, Franz and Fuchs, Paul and Zavadlav, Julija},
  journal={arXiv preprint arXiv:2510.11148},
  year={2025}
}

@article{yap2008coarse,
  title={A coarse-grained $\alpha$-carbon protein model with anisotropic hydrogen-bonding},
  author={Yap, Eng-Hui and Fawzi, Nicolas Lux and Head-Gordon, Teresa},
  journal={Proteins: Structure, Function, and Bioinformatics},
  volume={70},
  number={3},
  pages={626--638},
  year={2008},
  publisher={Wiley Online Library}
}

@article{diggins2018optimal,
  title={Optimal coarse-grained site selection in elastic network models of biomolecules},
  author={Diggins IV, Patrick and Liu, Changjiang and Deserno, Markus and Potestio, Raffaello},
  journal={Journal of chemical theory and computation},
  volume={15},
  number={1},
  pages={648--664},
  year={2018},
  publisher={ACS Publications}
}

@article{tozzini2006mapping,
  title={Mapping all-atom models onto one-bead coarse-grained models: general properties and applications to a minimal polypeptide model},
  author={Tozzini, Valentina and Rocchia, Walter and McCammon, J Andrew},
  journal={Journal of chemical theory and computation},
  volume={2},
  number={3},
  pages={667--673},
  year={2006},
  publisher={ACS Publications}
}

@article{mioduszewski2023contact,
  title={Contact-based molecular dynamics of structured and disordered proteins in a coarse-grained model: Fixed contacts, switchable contacts and those described by pseudo-improper-dihedral angles},
  author={Mioduszewski, {\L}ukasz and Bednarz, Jakub and Chwastyk, Mateusz and Cieplak, Marek},
  journal={Computer Physics Communications},
  volume={284},
  pages={108611},
  year={2023},
  publisher={Elsevier}
}

@article{kock2008growth,
  title={Growth behavior and intrinsic properties of vapor-deposited iron palladium thin films},
  author={Kock, Iris and Edler, T and Mayr, Stefan G},
  journal={Journal of Applied Physics},
  volume={103},
  number={4},
  year={2008},
  publisher={AIP Publishing}
}

@article{ercolessi1994interatomic,
  title={Interatomic potentials from first-principles calculations: the force-matching method},
  author={Ercolessi, Furio and Adams, James B},
  journal={Europhysics Letters},
  volume={26},
  number={8},
  pages={583},
  year={1994},
  publisher={IOP Publishing}
}

@article{shell2008relative,
  title={The relative entropy is fundamental to multiscale and inverse thermodynamic problems},
  author={Shell, M Scott},
  journal={The Journal of chemical physics},
  volume={129},
  number={14},
  year={2008},
  publisher={AIP Publishing}
}

@article{reith2003deriving,
  title={Deriving effective mesoscale potentials from atomistic simulations},
  author={Reith, Dirk and P{\"u}tz, Mathias and M{\"u}ller-Plathe, Florian},
  journal={Journal of computational chemistry},
  volume={24},
  number={13},
  pages={1624--1636},
  year={2003},
  publisher={Wiley Online Library}
}

@article{white2015designing,
  title={Designing free energy surfaces that match experimental data with metadynamics},
  author={White, Andrew D and Dama, James F and Voth, Gregory A},
  journal={Journal of Chemical Theory and Computation},
  volume={11},
  number={6},
  pages={2451--2460},
  year={2015},
  publisher={ACS Publications}
}

@article{adams2021learning,
  title={Learning 3d representations of molecular chirality with invariance to bond rotations},
  author={Adams, Keir and Pattanaik, Lagnajit and Coley, Connor W},
  journal={arXiv preprint arXiv:2110.04383},
  year={2021}
}

@article{das2024amino,
  title={Amino acid chirality: stereospecific conversion and physiological implications},
  author={Das, Kuladeep and Balaram, Hemalatha and Sanyal, Kaustuv},
  journal={ACS omega},
  volume={9},
  number={5},
  pages={5084--5099},
  year={2024},
  publisher={ACS Publications}
}

@article{ballard2020racemisation,
  title={Racemisation in chemistry and biology},
  author={Ballard, Andrew and Narduolo, Stefania and Ahmed, Hiwa O and Keymer, Nathaniel I and Asaad, Nabil and Cosgrove, David A and Buurma, Niklaas J and Leach, Andrew G},
  journal={Chemistry--A European Journal},
  volume={26},
  number={17},
  pages={3661--3687},
  year={2020},
  publisher={Wiley Online Library}
}

@article{barducci2008well,
  title={Well-tempered metadynamics: a smoothly converging and tunable free-energy method},
  author={Barducci, Alessandro and Bussi, Giovanni and Parrinello, Michele},
  journal={Physical review letters},
  volume={100},
  number={2},
  pages={020603},
  year={2008},
  publisher={APS}
}

@article{Laio2002,
    title = {{Escaping free-energy minima}},
    year = {2002},
    journal = {Proceedings of the National Academy of Sciences},
    author = {Laio, Alessandro and Parrinello, Michele},
    number = {20},
    pages = {12562--12566},
    volume = {99},
    publisher = {National Acad Sciences}
}

@article{schutt2018schnet,
  title={Schnet--a deep learning architecture for molecules and materials},
  author={Sch{\"u}tt, Kristof T and Sauceda, Huziel E and Kindermans, P-J and Tkatchenko, Alexandre and M{\"u}ller, K-R},
  journal={The Journal of chemical physics},
  volume={148},
  number={24},
  year={2018},
  publisher={AIP Publishing}
}

@article{gasteiger2020directional,
  title={Directional message passing for molecular graphs},
  author={Gasteiger, Johannes and Gro{\ss}, Janek and G{\"u}nnemann, Stephan},
  journal={arXiv preprint arXiv:2003.03123},
  year={2020}
}

@article{wilson2023anisotropic,
  title={Anisotropic molecular coarse-graining by force and torque matching with neural networks},
  author={Wilson, Marltan O and Huang, David M},
  journal={The Journal of Chemical Physics},
  volume={159},
  number={2},
  year={2023},
  publisher={AIP Publishing}
}

@article{unke2021machine,
  title={Machine learning force fields},
  author={Unke, Oliver T and Chmiela, Stefan and Sauceda, Huziel E and Gastegger, Michael and Poltavsky, Igor and Schutt, Kristof T and Tkatchenko, Alexandre and Muller, Klaus-Robert},
  journal={Chemical Reviews},
  volume={121},
  number={16},
  pages={10142--10186},
  year={2021},
  publisher={ACS Publications}
}

@article{behler2016perspective,
  title={Perspective: Machine learning potentials for atomistic simulations},
  author={Behler, J{\"o}rg},
  journal={The Journal of chemical physics},
  volume={145},
  number={17},
  year={2016},
  publisher={AIP Publishing}
}

@article{noe2020machine,
  title={Machine learning for molecular simulation},
  author={No{\'e}, Frank and Tkatchenko, Alexandre and M{\"u}ller, Klaus-Robert and Clementi, Cecilia},
  journal={Annual review of physical chemistry},
  volume={71},
  number={1},
  pages={361--390},
  year={2020},
  publisher={Annual Reviews}
}

@inproceedings{gilmer2017neural,
  title={Neural message passing for quantum chemistry},
  author={Gilmer, Justin and Schoenholz, Samuel S and Riley, Patrick F and Vinyals, Oriol and Dahl, George E},
  booktitle={International conference on machine learning},
  pages={1263--1272},
  year={2017},
  organization={Pmlr}
}


\end{document}